\documentclass[11pt,twoside]{article}
\usepackage{asp2014}

\aspSuppressVolSlug
\resetcounters

\begin{document}

\title{Advanced Data Analysis for Observational Cosmology: applications to the study of the Intergalactic Medium}

% full name: Guido Cupani
\author{Guido~Cupani$^{1,2}$, Giorgio~Calderone$^1$, Stefano~Cristiani$^{1,2}$, and Francesco~Guarneri$^1$}
\affil{$^1$INAF--Astronomical Observatory of Trieste, I-34151 Trieste, Italy; \email{guido.cupani@inaf.it}}
\affil{$^2$IFPU--Institute for Fundamental Physics of the Universe, via Beirut 2, I-34151 Trieste, Italy}
% remove/add as you need

% remove/add authors as you need
\paperauthor{Guido~Cupani}{guido.cupani@inaf.it}{0000-0002-6830-9093}{INAF}{Astronomical Observatory of Trieste}{Trieste}{Trieste}{I-34143}{Italy}
\paperauthor{Giorgio~Calderone}{giorgio.calderone@inaf.it}{0000-0002-7738-5389}{INAF}{Astronomical Observatory of Trieste}{Trieste}{Trieste}{I-34143}{Italy}
\paperauthor{Stefano~Cristiani}{stefano.cristiani@inaf.it}{0000-0002-2115-5234}{INAF}{Astronomical Observatory of Trieste}{Trieste}{Trieste}{I-34143}{Italy}
\paperauthor{Francesco~Guarneri}{francesco.guarneri@inaf.it}{}{INAF}{Astronomical Observatory of Trieste}{Trieste}{Trieste}{I-34143}{Italy}
% remove/add as you need

% leave these next few aindex lines commented for the editors to enable them. Use Aindex.py to generate them for yourself.
% first presenting author should be the first entry for bold-facing the author index page-reference
%\aindex{Cupani,~G.}
%\aindex{Author2,~S.}
% remove/add as you need

% leave the ssindex lines commented for the editors to enable them, use Index.py to suggest yours
%\ssindex{astronomy!quasar}
%\ssindex{data!analysis!spectral}
%\ssindex{software!spectral analysis}
%\ssindex{databases!MariaDB}
%\ssindex{techniques!Random Forest}

% leave the ooindex lines commented for the editors to enable them, use ascl.py to suggest yours
%\ooindex{QSFit, ascl:1612.011} 
%\ooindex{GAIA, ascl:1403.024} 
%\ooindex{Spectra, ascl:1701.003} 
%\ooindex{PRF, ascl:1903.009} 
  
\begin{abstract}
The analysis of absorption features along the line of sight to distant sources is an invaluable tool for observational cosmology, giving a direct insight into the physical and chemical state of the inter/circumgalactic medium.
Such endeavour entails the accessibility of bright QSOs as background beacons,
and the availability of software tools to %properly analyze the data,
extract the information in a reproducible way. In this article, we will present the latest results we obtained in both
directions within the QUBRICS project: 
we will describe how machine learning techniques 
were applied to detect hundreds of previously unknown QSOs in the southern hemisphere, and how state-of-the art software like QSFit and Astrocook was integrated in the analysis of the targets, opening up new possibilities for the next era of %intergalactic medium 
observations.
  
\end{abstract}

\section{Introduction: the quest for bright beacons}

Several science cases in cosmology and fundamental physics rely on the availability of luminous background beacons (typically quasars, or QSOs) at high redshift, to shed light on the intervening matter. Here is a partial list (see \citealt{2019-Calderone_QUBRICS} for details and references): (i) the determination the matter power spectrum at small scales; (ii) the measure of the abundances of primordial elements; (iii) the measure of a possible variation of fundamental constants; (iv) the Sandage Test, i.e. the direct measurement of the cosmic expansion rate from the redshift drifts of distant objects. 

Historically, there has been a dearth of detected luminous QSOs in the Southern hemisphere with respect to the North, due to the lack of appropriate surveys like the SDSS \citep{2017-Blanton_SDSS}. The scenario is now changing, thanks to several recent photometric databases (see Section \ref{sec:db}). It is therefore crucial to (i) mine the databases to identify quasar candidates; (ii) acquire the spectra of selected candidates to refine the mining techniques; (iii) consolidate the confirmed QSOs into a database; (iv) develop and test the analysis procedures required to pursue the science cases described above. The QUBRICS project (QUasars as BRIght beacons for Cosmology in the Southern hemisphere, \citealt{2019-Calderone_QUBRICS, 2020-Boutsia_QUBRICS, 2021-Boutsia_QUBRICS}) is addressing all these tasks: it has identified so far several hundreds of new QSOs at $\delta<0$, which are now suitable to be observed with ground-based facilities like VLT UVES and VLT ESPRESSO, and in the future with the ESO ELT. 

\section{The QUBRICS database}\label{sec:db}

%The QUBRICS project aims to identify new, bright and high-redshift QSOs (or beacons) in the Southern Hemisphere \citep{2019ApJ...887..268C}. We used photometric data in publicly available surveys, analyzed by means of machine learning methods in order to identify QSO candidates to be observed spectroscopically using facilities in the South. Finally, we used the acquired knowledge to address specific aspects of cosmology.  The QUBRICS project already produced several published papers, a few more are currently under review, and more papers are yet to come.

The problem of detecting high-$z$ QSOs in the sky entails two different tasks: (i) distinguishing QSOs from stars, galaxies, and other sources; (ii)
estimating the redshift of the source to discard low-$z$ candidates. To tackle the problem, we collected photometric data from various sources into a database, and used it to train both classification and regression models, in order to identify QSO candidates.

Photometric data from the Skymapper \citep{2018-Wolf_SkyMapperDR1} and the PanSTARRS \citep{2018-Chambers_PanSTARRS} surveys were cross-matched %them
against the WISE \citep{2010-Wright_WISE}, GAIA \citep{2016-GaiaDR1}, and 2MASS \citep{2006-Skrutskie_2MASS} catalogs. Matching was restricted to targets with $\delta<15\deg$, galactic latitude $|b|>25\deg$, and a limiting magnitude $i<18$ ($Y<18.5$ since 2021). Ambiguous matches were flagged and rejected. The resulting entries are ``standardized'' (e.g. translating their magnitudes to the same photometric system) and fed into a single database, where they are assigned a unique ``QUBRICS ID'' (\emph{qid} for short). The database is hosted on a 4TB (RAID01) machine and managed with MariaDB\footnote{https://mariadb.org} using Julia \citep{2017-Bezanson_Julia} scripts for maintenance. The overall architecture of the database is summarized in Figure \ref{fig:db}. 
\articlefigure[width=0.8\textwidth]{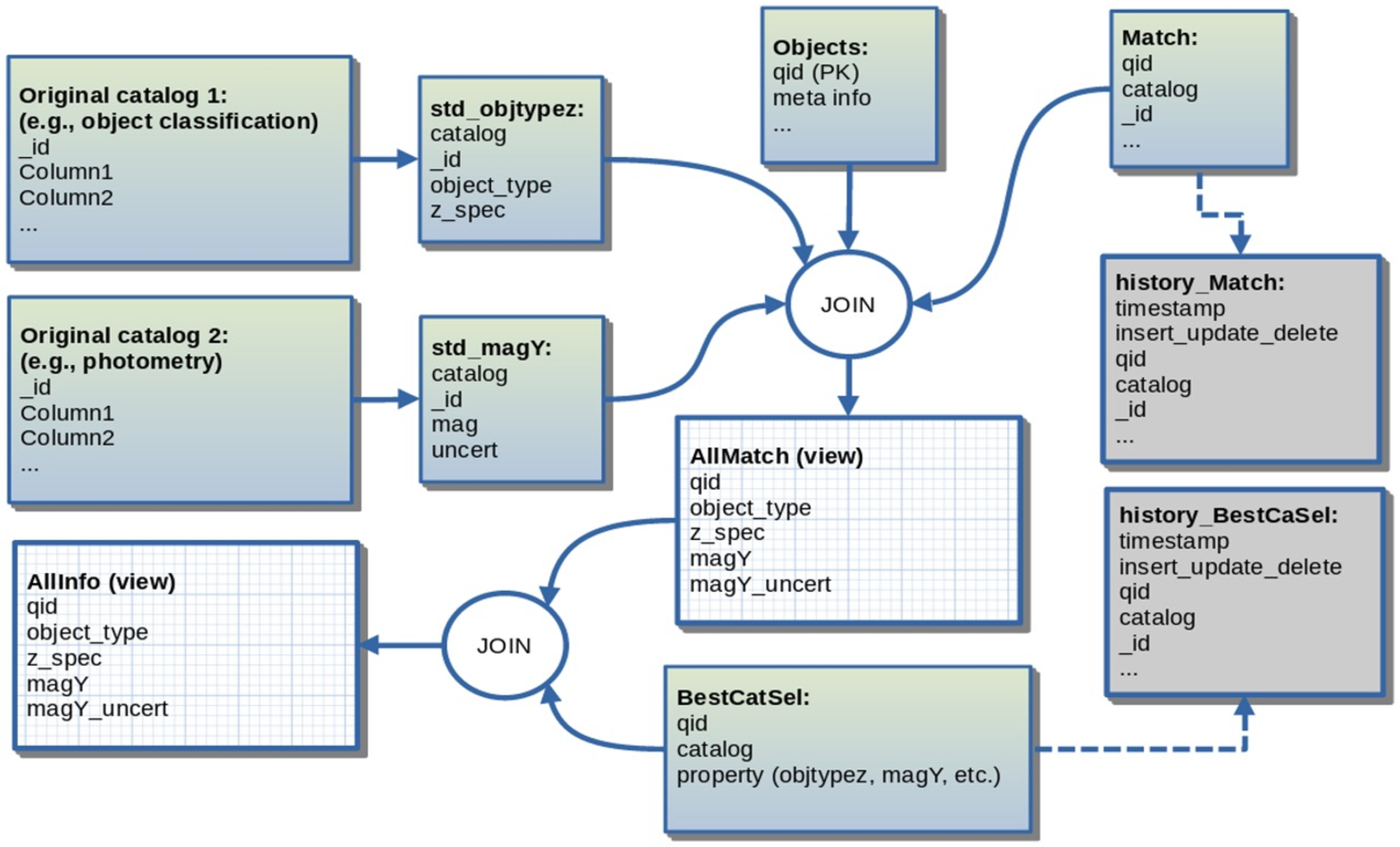}{fig:db}{Simplified diagram of the database architecture.}
In the database, the pieces of information information (including the correspondence between the \emph{qid} and the source catalogs with their original identifiers) are joined together produce a ``live view'' with all columns from the different catalogs, and a row for each match ({\it AllMatch}). The history of the modifications is also preserved. 

The selection of QSO candidates is performed on a filtered version of the ``live view''. For each \emph{qid} and property, the {\it BestCatSel} table reports the catalog that provides the most reliable value, resulting in a reduced view with a single single row for each \emph{qid} ({\it AllInfo}). All entries in this view can always be traced back to the original catalog using the identifiers. Note that if a more reliable catalog is added, the new values can be uploaded by simply updating the filtering table.

%Another important feature provided by the DB system is that it stores all history of modifications of the most relevant tables, so that it is always possible to analyze the evolution of the values for each source.  The whole DB takes advantage of the triggering mechanisms to automatically populate the history tables, as well as to keep the cache table (not shown here) up to date.

\section{Candidate selection and confirmation} 

The database described in the previous section was used for training different selection techniques to detect QSO candidates. The first QUBRICS selection was based on the Canonical Correlation Analysis (CCA), a higher dimensional selection process based on optimized linear combinations of photometric colors \citep{2019-Calderone_QUBRICS}. Since then, more advanced techniques have been applied, such as the Probabilistic Random Forest (PRF) and, recently, the XGBoost algorithm \citep{2016-Chen_XGBoost}. 

The PRF \citep{2021-Guarneri_QUBRICS} is a modification of the original Random Forest algorithm, designed to properly handle uncertainties by representing the features of the input data as probability distribution functions. XGB, on the other hand, uses gradient boosting to accommodate model misclassification while keeping overfitting under control. Both algorithms are trained to perform the selection in two stages: (i) discriminating QSOs from stars and galaxies; (ii) rejecting sources at with $z<2.5$. This second task is not trivial, as the training dataset is heavily skewed toward low-$z$ objects (a feature which was accounted for through oversampling techniques). Despite the difficulties, both algorithms proved effective in terms of precision and recall. XGB was also used to estimate the redshift of the candidates throgut regression. 

A spectroscopic follow-up campaign was carried out to confirm the detected candidates. Around 350 new high-$z$ QSOs have been confirmed so far (see Figure \ref{fig:new}). Synthetic data are currently being tested to improve the performance of the algorithms.

\articlefigure[width=0.7\textwidth]{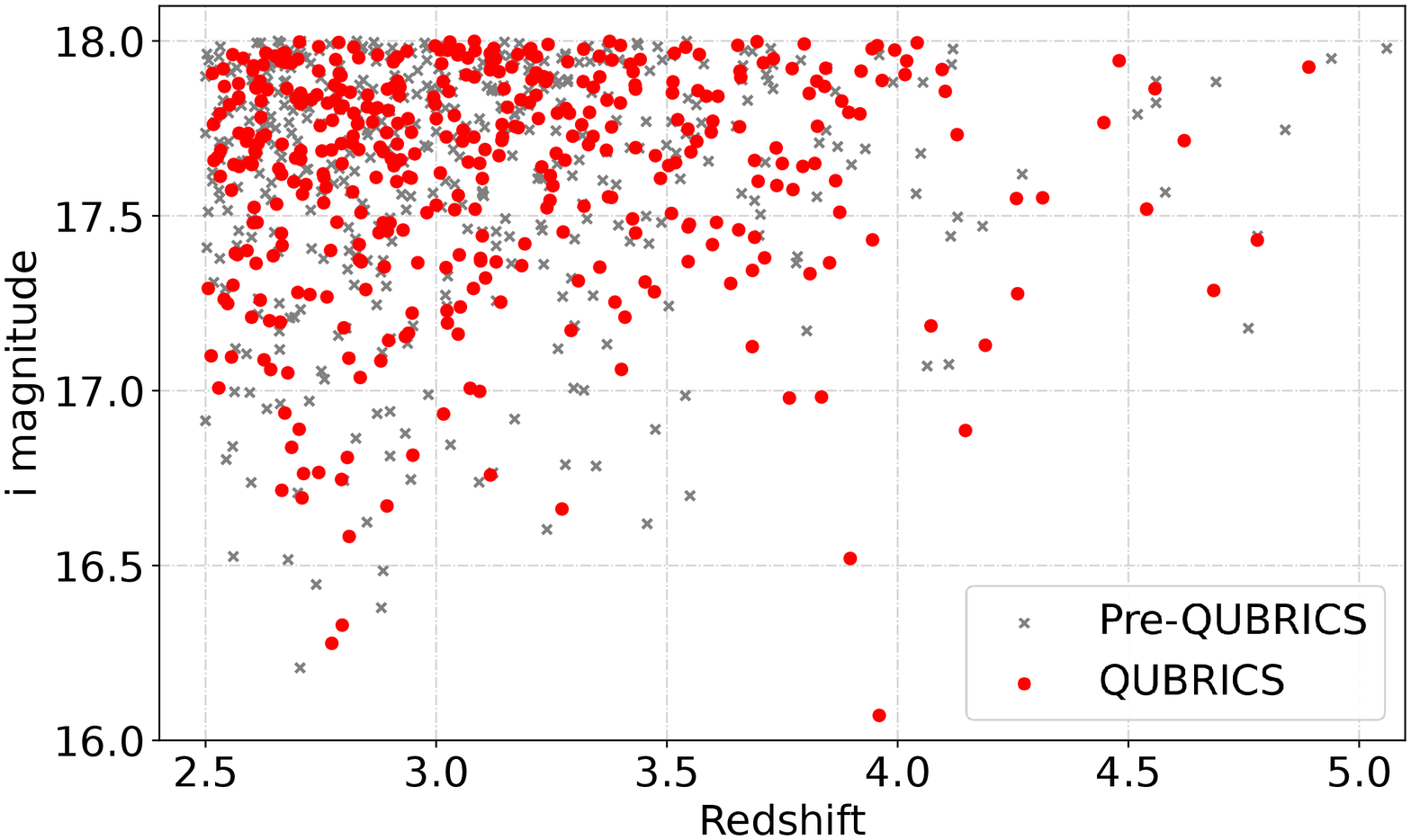}{fig:new}{New high-$z$ quasars confirmed by QUBRICS (red dots), compared with those already known from literature (black dots).}

\section{Data analysis: a test case for Astrocook} 

A natural complement to the results obtained by QUBRICS is the development of workflows to analyze the collected data. These workflows must not only be scientifically reliable, but also durable (i.e. reproducible and easily maintained). Stability of the data analysis through some decades is essential, for example, to perform the Sandage test \citep{2008-Liske_ST}; more generally, the reproducibility paradigm is getting traction in several fields, as a way to cope with increasingly complex science cases. 

These requirements laid the foundation for Astrocook \citep{2020-Cupani_Astrocook}, a Python package to design, run, and share analysis workflows for QSO spectra. Within the ``cooking'' metaphor, Astrocook provides both a set of ``recipes'' (procedures to manipulate spectral data and model emission and absorption features) and a ``kitchen'' (including a graphical user interface and a scripting-logging tool) to let the user prepare their own ``dishes'' in a controlled, repeatable way. The scripting/logging tool, in particular, can be used at run-time to transform workflows into readable JSON files, serving both as a documentation and as a way to reproduce the analysis at a later time. 

Within the QUBRICS project, Astrocook was used in combination with another tool, QSFit \citep{2019-Calderone_QSFit}, to analyze a peculiar class of objects that emerged from the survey. These objects were initially selected as high-$z$ candidates, but could not be confirmed as bona-fide QSOs from optical spectroscopy alone. Most of them were identified through near-infrared spectroscopy as broad absorption line (BAL) QSOs, with a notable fraction showing strong Fe\textsc{ii} absorption bluewards from the emission redshift (FeLoBAL QSOs). The analysis of these objects is currently in press; notably, we are going to publish it together with the complete set of procedures to re-run it, to enforce the reproducibility paradigm and to foster further investigation. 

\section{Conclusions}

We have described the QUBRICS project in the context of the rapid evolution of cosmology into a precision science. We have shown how the most fundamental scientific questions in this fields can be addressed only with a synergistic approach, combining several state-of-the-art software solutions to cover all the steps from data mining to data interpretation. This is our effort and our challenge as we look forward to the next generation of extremely large telescopes.

% For example in \citet{PID_adassxxx} it was shown that ...

\bibliography{O1-003}

% if we have space left, we might add a conference photograph here. Leave commented for now.
% \bookpartphoto[width=1.0\textwidth]{foobar.eps}{FooBar Photo (Photo: Any Photographer)}

\end{document}